\documentclass{aa}
\usepackage{natbib}
\usepackage{graphicx}
\bibpunct{(}{)}{;}{a}{}{,} 

\begin{document}

\title{Jets, accretion, coronae and all that: The enigmatic X-rays from the Herbig star HD 163296 \thanks{Based on observations obtained with XMM-Newton, an ESA science mission with instruments and contributions directly funded by ESA Member States and NASA.}\thanks{Fig.~\ref{rgs} is available in electronic form via http://www.edpsciences.org}}

\titlerunning{X-ray emission from HD 13296}

\author{H.~M. G\"unther \and J.~H.~M.~M. Schmitt}  
\institute{Hamburger Sternwarte, Universit\"at Hamburg, Gojenbergsweg 112, 21029 Hamburg, Germany\\ \email{moritz.guenther@hs.uni-hamburg.de}}
\date{Received 22 September 2008 / accepted 28 November 2008}
\abstract{Herbig Ae/Be stars (HAeBe) are pre-main sequence objects in the mass range 2~M$_{\odot} < M_* < $~8~M$_{\odot}$. Their X-ray properties are uncertain and, as yet, unexplained.}{We want to elucidate the X-ray generating mechanism in HAeBes.}{We present a \emph{XMM-Newton} observation of the HAeBe HD~163296. We analyse the light curve, the broad band and the grating spectra, fit emission measures and abundances and apply models for accretion and wind shocks.}{We find three temperature components ranging from 0.2~keV to 2.7~keV. The \ion{O}{vii} He-like triplet indicates a X-ray formation region in a low density environment with a weak UV photon field, i.~e. above the stellar surface. This makes an origin in an accretion shock unlikely, instead we suggest a shock at the base of the jet for the soft component and a coronal origin for the hot component. A mass outflow of $\dot M_{\rm shock} \approx 10^{-10} M_{\sun}$~yr$^{-1}$ is sufficient to power the soft X-rays.}{ HD~163296 is thought to be single, so this data represent genuine HAeBe X-ray emission. HD~163296 might be prototypical for its class.}
\keywords{stars: formation -- stars: individual: HD 163296 -- X-rays: stars}

\maketitle

\section{Introduction}
\label{introduction}
The phase between protostar and main-sequence object is a key stage for planet formation and, furthermore, the properties of a star are determined precisely during that phase for the rest of its life. It is therefore important to study objects of this age in order to understand the origin of planetary systems. Herbig Ae/Be stars (HAeBes) are thought to be the predecessors of main-sequence (MS) stars in the mass range 2~M$_{\odot} < M_* < $~8~M$_{\odot}$, although the empirically defined object class of HAeBes may in fact encompass a wider range of evolutionary stages such as stars with magnetospheric accretion (Ae) and boundary layer objects (Be). HAeBes are certainly young, they do have clear signatures of surrounding disks and thus must be considered to be the more massive brothers of the better studied classical T Tauri stars (CTTS), which are young ($<10\;\mathrm{Myr}$), low mass ($M_*<3M_{\odot}$), pre-main sequence stars exhibiting strong H$\alpha$ emission. CTTS are also surrounded by disks, as evidenced by a strong infrared excess, and are actively accreting material.

Strong X-ray emission is a characteristic property of most young stellar objects. A review of the observational situation before \emph{XMM-Newton} and \emph{Chandra} for low mass stars is given by \citet{1999ARA&A..37..363F}. \citet{1994A&A...292..152Z} were the first to carry out a systematic X-ray survey of HAeBes. Restricting attention to those stars located within 500~pc, 70~\% of all HAeBes were found to produce detectable X-ray emission with X-ray luminosities ranging between a few $10^{29}$~erg~s$^{-1}$ and $10^{31}$~erg~s$^{-1}$, thus exhibiting a much larger detection rate than found for field A-type stars (10-15~\%, according to \citet{2007A&A...475..677S}).
Also, \citet{2004A&A...413..669G} report the detection of a large X-ray flare (with \emph{XMM-Newton}) from the Herbig Ae star \object{V892 Tau}, and present strong arguments that the flare in fact originated from the Ae star in the V892 Tau system; note, however, that \citet{2005A&A...431..307S} show V892~Tau to be a close binary. Until today it is unclear whether the observed X-ray emission originates from the HAeBes themselves or from unresolved companions, which by necessity would have to be low-mass, young, and active stars. \citet{2004ApJ...614..221S} studied a sample of ten close HAeBes and showed that the X-ray emission probably originates from a magnetically confined plasma, although it remains unclear if this plasma is associated with a companion. In a sample of 17~HAeBes  \citet{2006A&A...457..223S} find X-ray emission in about 80~\% of the objects, more than can be reasonably expected from late-type companions, although it is possible that their sample is biased towards known X-ray sources. In a follow-up paper \citet{BeateHAeBe} detect every object, but again many of them show spectral characteristics compatible with low-mass companions.

X-ray astronomy has been revolutionised by the availability of high resolution grating spectroscopy in the current generation of X-ray satellites. Grating spectra of HAeBes published so far are taken from \object{AB Aur} \citep{ABAur} and the spectroscopic binary \object{HD 104237} \citep{2008ApJ...687..579T}, where the main component is a Herbig~Ae star with a CTTS companion of spectral type K3. More observations are available from CTTS; their high resolution X-ray spectra typically show a soft component (which may be hidden by large absorption columns) and  unusually low f/i-ratios in the He-like triplets of Ne and O. For CTTS the most promising explanation for these phenomena is a magnetically funnelled infall model. The inner disk is truncated at the corotation radius and ionised material is loaded onto the field, flowing along the magnetic field lines and hitting the stellar surface close to free-fall velocities \citep[e.g.][]{1994ApJ...429..781S}. In the photosphere an accretion shock develops, which produces the soft X-ray component. Because the plasma has relatively high densities, the f/i-ratios in the \ion{Ne}{ix} and \ion{O}{vii} triplets are small. This model has been successfully applied in explaining the X-ray emission from the CTTS \object{TW Hya} in terms of an accretion shock model plus a hot corona \citep{acc_model}.

A question unanswered to date is whether the accretion disks surrounding HAeBes lead to similar phenomena as those surrounding CTTS or if other modes of X-ray generation operate in the more massive stars. This could be, e.g., magnetically confined winds  colliding in the equatorial plane as suggested for \object{IQ Aur} by \citet{1997A&A...323..121B}, or internal shocks in unstable winds as in the CTTS \object{DG Tau} \citep{2005ApJ...626L..53G,2008A&A...478..797G,Schneider,dgtau}.

All these models require magnetic fields, which are only weak in evolved A and B stars, because of the absence of an outer convection zone required for a solar-type $\alpha-\Omega-$dynamo. Nevertheless the HAeBes could have magnetic fields, produced by the compression of a primordial field of the proto-stellar cloud. It has been suggested that they are the progenitors of the magnetic Ap/Bp stars \citep{2005A&A...442L..31W}, which comprise about 5\% of the total A star population. 

In order to enlighten the origin of the X-rays from HAeBes we performed X-ray observations of \object{HD 163296} with \emph{XMM-Newton}, whose stellar properties we explain in Sect.~\ref{stellarproperties} before presenting the observations in Sect.~\ref{observations}. We show the results of our analysis in Sect.~\ref{results} and discuss their implications in Sect.~\ref{discussion}. Sect.~\ref{conclusion} gives our conclusions.

\section{Stellar properties}
\label{stellarproperties}
HD~163296 is an isolated HAeBe far from any natal molecular cloud. With its distance of $122^{+17}_{-13}$~pc \citep{1998A&A...330..145V} it is one of the closest objects of its kind, making it an ideal target for studies of young A stars. The presence of cold dust around HD~163296 has long been known from spatially unresolved observations \citep{1994MNRAS.271..587M}, and a few years ago the disk of HD 163296 has been coronographically imaged with \emph{HST}/STIS showing indications for a planetary body in the disk \citep{2000ApJ...544..895G}. Furthermore, these observations allow to trace an outflow in the Ly$\alpha$ line from 7.3~AU to 725~AU with a velocity of $\approx350$~km~s$^{-1}$ \citep{2000ApJ...542L.115D}. This jet, called \object{HH 409}, contains knots up to about 3000~AU from the central star, with an asymmetry between jet and counterjet \citep{2006ApJ...650..985W}. The surrounding disk has also been imaged in the millimetre range indicating a strongly evolved disk, where larger bodies already influence the evolution of the gas \citep{2007A&A...469..213I}. The spectrum of HD~163296 displays a strong infrared excess, which can be attributed to the inner disk regions. Apart from a single outburst the optical lightcurve has been remarkably constant over the last 25 years \citep{2008ApJ...678.1070S}.

The star itself is of spectral type A1 with an effective temperature $T_{\mathrm{eff}}=9300$~K, a radius $R=2.1\;R_{\sun}$ and a mass $M=2.3\; M_{\sun}$. According to evolutionary models this places it at an age of about 4~Myr. With an $A_V=0.25$ the star is only mildly absorbed \citep{1998A&A...330..145V}. 
\citet{2006A&A...446.1089H} performed spectropolarimetric observations of circularly polarised light at the VLT for some HAeBes, but were unable to detect a significant longitudinal magnetic field ($B_Z=-57\pm33$~G) on HD~163296; they remark, however, that a monitoring campain, which could detect accretion funnels, is still missing. 
\citet{2005A&A...429..247D} present \emph{Far ultraviolet spectroscopic explorer (FUSE)} and \emph{HST}/STIS observations of HD~163296 showing chromospheric signatures and over-ionised species. These can be explained by either an accretion shock or in a magnetically confined wind model. Hydrogen Ly$\alpha$ and \ion{C}{iii} lines show outflow signatures.
\citet{2005ApJ...628..811S} looked for signatures of a companion, but \emph{HST}/STIS imagery excludes binarity down to 0\farcs05 separation. Their long-slit STIS spectra also show only the emission of an early-type unresolved point source. 

HD 163296 is a \emph{ROSAT} X-ray source, detected in the \emph{ROSAT} all-sky survey (RASS) data with an elongated source near the position of HD 163296 and resolved in three sources with an unpublished \emph{ROSAT}/HRI pointing. \citet{2005ApJ...628..811S} confirm the detection of X-ray emission from  HD~163296 using {\it Chandra} ACIS-S in imaging mode. They find an extremely soft spectrum which can be well fitted with a 1-temperature model (k$T\approx 0.5$~keV, where k is Boltzmann's constant and $T$ is the temperature). Additionally there are five X-ray photons at the position of the knot H in the jet. This is a significant source detection at the 95~\% confidence level, with a luminosity more than two orders of magnitude below the central component.

\section{Observations and data reduction}
\label{observations}
We observed HD~163296 for 130~ks with \emph{XMM-Newton} with the RGS as prime instrument applying the medium filter to block out the bright optical radiation. The observation is split in two exposures in consecutive orbits. Additionally we retrieved  archival observations from \emph{XMM-Newton} and \emph{Chandra}. Table~\ref{obslog} summarises the observation information.
\begin{table}
\caption{\label{obslog}Observing log.}
\begin{center}
\begin{tabular}{lllr}
\hline \hline
Observatory & ObsID  & Obs. date & Exp. time \\
\hline
Chandra & 3733 & 2003-08-10 & 20 ks\\
XMM-Newton & 0144271401 & 2003-10-11 & 7 ks\\
XMM-Newton & 0502370201 & 2007-09-23 & 106 ks\\
XMM-Newton & 0502370301 & 2007-09-24 & 22 ks\\
\hline
\end{tabular}
\end{center}
\end{table}
HD~163296 is optically bright, just below the formal brightness limits of the UVW2 and UVM2 filters, but all exposures of the optical monitor are overexposed and unusable for analysis in the inner region. All data was reduced using standard \emph{XMM-Newton} Science Analysis System (SAS) software, version 7.1 or the \emph{Chandra} Interactive Analysis of Observations (CIAO) software, version 4.0, in the case of the \emph{Chandra} observation. Because the exposure is partially contaminated by energetic proton events we applied a time filtering, where a good time interval is defined in the usual way for the EPIC/PN camera as $<1$~cts~s$^{-1}$. For the EPIC/MOS detectors we lowered the cut values to 0.25~cts~s$^{-1}$ in order to suppress the high energy noise in the spectra. CCD-spectra and count rates were extracted from a circular region within the 15\arcsec{} around the target for the MOS. We obtained RGS spectra and found a strong contamination in the RGS1. We therefore reduced the extraction region for the RGS1 to the central 66\% of the point spread function (PSF), keeping 90\% of the PSF extraction region for the RGS2. This step helps to suppress the background.  To increase the signal we merged the two exposures for each RGS using the SAS task \texttt{rgscombine}. Spectral fitting was carried out using XSPEC V11.3 \citep{1996ASPC..101...17A}, and individual line fluxes were measured using the CORA line fitting tool \citep{2002AN....323..129N}. Because the line widths are dominated by instrumental broadening we keep them fixed at $\Delta\lambda=0.06$~\AA{}.

For comparison purposes we also analysed the \emph{XMM-Newton} data on AB~Aur (ObsID 0101440801) which was already presented in great detail by \citet{ABAur}. We extracted the central 80\% of the PSF in order to minimise the contribution from the close source \object{SU Aur} and measured the line fluxes in the He-like triplets of neon and oxygen with the same method used for HD~163296.

\section{Results}
\label{results}
We first present the variability observed in HD~163296, before we show fits to the emission measure distribution. Then we give individual line fluxes and take a closer look at the density and UV-field sensitive He-like triplets. We compare the \ion{O}{viii} to \ion{O}{vii} line ratio with the value found for MS stars.

\subsection{Variability}
\begin{figure}
\resizebox{\hsize}{!}{\includegraphics{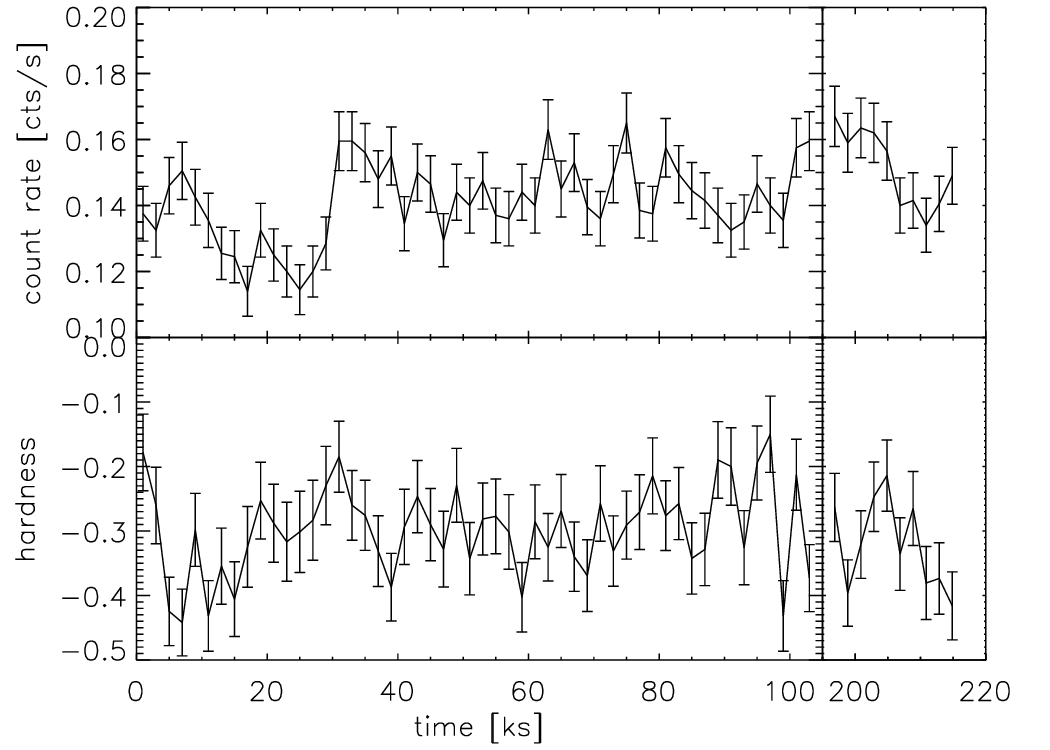}}
\caption{Lightcurve and hardness (soft band: 0.2-0.8~keV, hard band: 0.8-2.0~keV) during the observations in 2007.}
\label{lc}
\end{figure}

Figure~\ref{lc} shows the light curve and hardness ratio of the observations taken in 2007. They are binned to 2~ks. We define the hardness ratios as a count ratio of (hard-soft)/(hard+soft), where the hard band is taken in the range 0.8-2.0~keV and the soft band as 0.2-0.8~keV. HD~163296 exhibits only modest variability over most of the observations with a sudden increase of the count rate by 30\% about 30~ks in the observation. No correlation between the hardness and the luminosity is visible. In the beginning low count rates coincide with a hard spectrum, but this pattern does not persist. 

To analyse the long-term variability we compare the count rates in the EPIC/PN detectors of all \emph{XMM-Newton} observations. In 2003 the average count rate was $0.166\pm0.007$~cts~s$^{-1}$ and in 2007 $0.202\pm0.002$~cts~s$^{-1}$ and $0.218\pm0.004$~cts~s$^{-1}$. The errors are statistical and do not represent temporal variation within each exposure. The exposure in 2003 is too short to extract a meaningful lightcurve. We obtained a \emph{ROSAT}/HRI count rate of 0.012~cts~s$^{-1}$ for a pointing on HD~163296 observed in 1995 from the HRI catalogue and converted it to an energy flux in the band 0.3-2.0~keV using WebPIMMS assuming that a spectral model fitted to the \emph{XMM-Newton} data is applicable. Fitting the same model to the \emph{Chandra} CCD spectrum we also integrated the model flux in the same band as the other observations. The \emph{ROSAT}/HRI luminosity is about $4.3\times10^{29}$~erg~s$^{-1}$, in 2003 \emph{Chandra} measured $3.8\times10^{29}$~erg~s$^{-1}$, with $4.1\times10^{29}$~erg~s$^{-1}$ the \emph{XMM-Newton} observation taken in the same year shows a flux very similar to the value from 1995. Our new observations are slightly brighter, with average fluxes of $4.5\times10^{29}$~erg~s$^{-1}$ and $4.8\times10^{29}$~erg~s$^{-1}$ respectively. The uncertainty of the luminosities is dominated by the intrinsic variation, not by count statistics. We find short-term variations of 30\% within a single exposure, and the separate observation differ on a similar scale. Thus, we conclude that the total luminosity has not changed much over the past decade in accordance with the optical behaviour \citep{2008ApJ...678.1070S}.

\subsection{Emission measure and abundance}\label{fitem}
\begin{figure}
\resizebox{\hsize}{!}{\includegraphics[angle=-90]{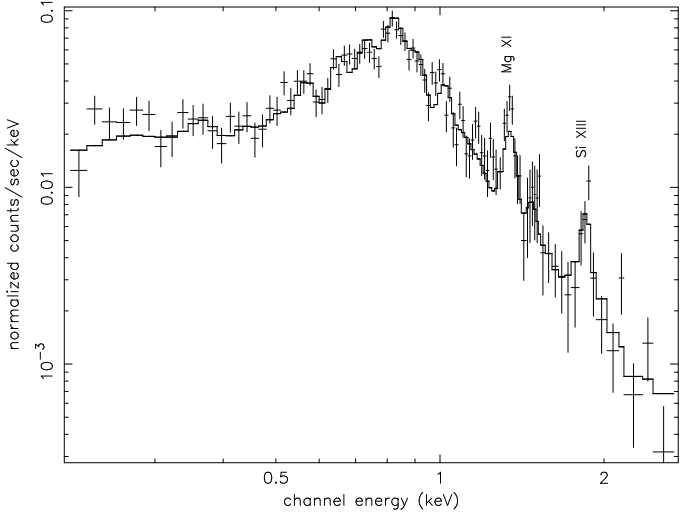}}
\caption{Low resolution spectrum (MOS1) of HD~163296 and our best-fit model.}
\label{mos}
\end{figure}

Because of the absence of significant variability during our observation in 2007 we jointly fit all available spectra from EPIC/MOS, keeping them separately, and the two RGS spectra, with the two exposures merged for each detector, again keeping the two detectors separately. The spectrum can be satisfactorily described by three thermal components with variable abundances. We use \texttt{VAPEC} models and give abundances relative to the solar values from \citet{1998SSRv...85..161G}. Our best fit model is shown in Fig.~\ref{mos}, where it is plotted as line on the MOS1 data. The data is binned to contain at least 15 counts per bin. We show the data only up to 3~keV because the source flux drops quickly and the high-energy spectrum is dominated by residual noise in the data.
The spectrum is nearly flat on the low energy side, indicating a weak absorption. Several emission peaks are visible, most notably the Ly$\alpha$ lines of \ion{Mg}{xii} at 1.35~keV and \ion{Si}{xiv} at 1.85~keV. Line contributions identified in the high resolution RGS spectra are listed in table~\ref{lineflux}. 

In tables~\ref{3vapecem} and \ref{3vapecab} we give the best fit parameters for our model and the associated errors (90\% confidence interval). The reduced $\chi^2$ value for our model is only 1.1, but fewer components cannot reproduce the data: A fit with two or only one component fails to produce the correct slope below 1~keV. This can only be compensated by extra  emission from unresolved carbon lines, which in turn requires a very low contribution from nitrogen around 0.5~keV. In total these models end up with a carbon abundance an order of magnitude above solar and virtually no nitrogen. We deem this an unphysical scenario and conclude that at least three temperature components are present in HD~163296. Within the error the fit using three components does not change if nitrogen is taken as a free parameter, but the error on its abundance is so high that we prefer to keep it fixed at the solar value.
As a cross-check we test the model with the data of the EPIC/PN detector, which has a larger effective area, but lower energy resolution; for this case we find the reduced $\chi_{\mathrm{red}}^2=1.1$.
\begin{table}
\caption{\label{3vapecem}Best-fit model parameters (90\% confidence interval).}
\begin{center}
\begin{tabular}{lllll}
\hline \hline
\multicolumn{2}{c}{component} &  soft & medium & hard \\
\hline
k$T$ & [keV] & $0.21^{+0.03}_{-0.01}$ & $0.51^{+0.1}_{-0.03}$ & $2.7^{+1.5}_{-0.8}$ \\
$EM$ & [$10^{52}$~cm$^{-3}$] & $2.3^{+0.8}_{-0.4}$ & $1.2^{+0.5}_{-0.6}$ & $0.5^{+0.1}_{-0.2}$\\
\hline\hline
$N_{\rm H}$ & [$10^{20}$~cm$^{-2}$] & \multicolumn{3}{c}{$7^{+4}_{-4}$}\\
\hline
\end{tabular}
\end{center}
\end{table}
\begin{table}\caption{\label{3vapecab}Abundance of elements and first ionisation potentials (FIP); errors show 90\% confidence intervals.}
\begin{center}
\begin{tabular}{ccc}
\hline\hline
Element & abundance & FIP [eV]\\
 \hline
C   & $3.7^{+3.6}_{-1.5}$ &11.3\\
O   & $0.7^{+0.2}_{-0.2}$ &13.6\\
Ne  & $1.2^{+0.5}_{-0.6}$ &21.6\\
Mg  & $2.3^{+1.0}_{-0.8}$ &7.6\\
Si  & $2.8^{+1.2}_{-0.8}$ &8.1\\
Fe  & $1.6^{+0.7}_{-0.4}$ &7.9\\
\hline
\end{tabular}
\end{center}
\end{table}

Table~\ref{3vapecab} shows the abundances as determined from the combined fit. Elements with a low first ionisation potential (FIP) are enhanced. A similar pattern is typical of inactive stars and its signatures can be found in many X-ray data sets \citep{2008A&A...486..995R}. 

We measure the absorbing column density towards the source to $N_{\mathrm{H}}=7_{-4}^{+4}\times 10^{20}$~cm$^{-3}$. Assuming a standard gas-to-dust ratio the optical reddening and the X-ray absorption should be related through the formula
$ N_{\mathrm{H}}=A_V \cdot2\times 10^{21}\mathrm{cm}^{-2}$ \citep[][see \citet{2003A&A...408..581V} for a compilation of other conversion factors in the literature, all roughly consistent with this value]{1979ARA&A..17...73S}. The optical reddening of $A_V=0.25$ \citep{1998A&A...330..145V} is fully consistent with the X-ray value.

\subsection{Line fluxes}
The spectrum is dominated by emission lines, it is shown binned to a minimum of five counts per bin in Fig.~\ref{rgs} (online only). Applying the global model to the RGS data alone gives $\chi_{\mathrm{red}}^2=1.7$, significantly more than in the joint fit. This is not surprising since our model of only three temperature components is a simplification of the real temperature distribution. Lines which are very sensitive to small temperature differences cannot match the model precisely. The fit can be improved to $\chi_{\mathrm{red}}^2=1.1$ by increasing the abundance of carbon by a factor of two and iron by a factor of four.
\onlfig{3}{
\begin{figure*}
\resizebox{\hsize}{!}{\includegraphics[angle=-90]{rgs05-10.ps}}
\resizebox{\hsize}{!}{\includegraphics[angle=-90]{rgs10-15.ps}}
\resizebox{\hsize}{!}{\includegraphics[angle=-90]{rgs15-20.ps}}
\resizebox{\hsize}{!}{\includegraphics[angle=-90]{rgs20-25.ps}}
\resizebox{\hsize}{!}{\includegraphics[angle=-90]{rgs25-30.ps}}
\resizebox{\hsize}{!}{\includegraphics[angle=-90]{rgs30-35.ps}}
\caption{High resolution RGS spectra with the best-fit model from the global fit in Sect.~\ref{fitem} with the corresponding fit residuals. The data is binned to a minimum of 5 counts per bin. {\it black:} RGS1 {\it red/grey:} RGS2}
\label{rgs}
\end{figure*}
}
As explained above we used a smaller extraction region for the RGS1. A comparison between spectra extracted using the standard extraction region and our reduced extraction region shows that our choice of extraction area reduces the noise level. The line fluxes measured with both methods are compatible within the errors. We list the line fluxes in table~\ref{lineflux}.
Due to zero-point offsets the measured wavelength and its theoretical value can differ; in the table we give the theoretical wavelength of the identified lines together with the fitted values. To fit multiplets we keep the wavelength difference between the components constant.
\begin{table*} [htb]
\caption{Measured line fluxes for HD~163296 with $1\sigma$ errors\label{lineflux}}
\begin{center}
\begin{tabular}{llrrrrrr}
\hline \hline
Line ID                 &$\lambda$ (theory)& \multicolumn{2}{c}{RGS1}        & \multicolumn{2}{c}{RGS2}       & flux        & unabs. intensity \\
                          & [\AA]          & $\lambda$ (fit)[\AA]& [counts]  &$\lambda$ (fit) [\AA]&[counts]  & [$10^{-6}$~cts~cm$^{-2}$~s$^{-1}$]& [$10^{27}$~erg~s$^{-1}$]    \\
\hline
\ion{Ne}{x} Ly\,$\alpha$  &12.14           & n.a.          &   n.a.          &$12.14\pm0.01$  &$ 38\pm 9$     &$5.1 \pm1.2 $&$ 17.7 \pm   6.2 $\\
\ion{Ne}{ix} r$^a$        &13.46           & n.a.          &   n.a.          &$13.45\pm0.01$  &$32 \pm 9  $   &$ 4.4\pm1.2 $&$ 14.5 \pm   5.7 $\\
\ion{Ne}{ix} i$^a$        &13.56           & n.a.          &    n.a.         &$13.55\pm0.01$  &$21\pm 9    $  &$ 2.9\pm1.2 $&$  9.6 \pm   4.8 $\\
\ion{Ne}{ix} f            &13.70           &   n.a.        &   n.a.          &$13.69\pm0.01$  &$22 \pm 8    $ &$ 3.0\pm1.1 $&$  9.9 \pm   4.6 $\\
\ion{O}{viii} Ly\,$\alpha$&18.97           &$18.98\pm0.01$ & $74 \pm 10   $  &$18.98\pm0.01$  &$117\pm13$     &$18.7\pm2.2 $&$ 54.6 \pm  20.4 $\\
\ion{O}{vii} r            &21.6            &$21.57\pm0.01$ & $42.7 \pm 7.8$  & n.a.           & n.a.          &$14.6\pm 2.7$&$ 45.2 \pm  21.5 $\\
\ion{O}{vii} i            &21.8            &$21.77\pm0.01$ & $ 7.8\pm 4.4$   & n.a.           & n.a.          &$ 3.2\pm1.8 $&$ 10.0 \pm   7.2 $\\
\ion{O}{vii} f            &22.1            &$22.07\pm0.01$ & $46.2  \pm 8.1$ & n.a.           & n.a.          &$14.6\pm2.6 $&$ 46.0 \pm  22.6 $\\
\ion{N}{vii} Ly\,$\alpha$ &24.78           & n.a.          & n.a.$^b   $     &$24.78\pm0.01$  &$14\pm 6$      &$3.3\pm 1.4 $&$  8.0 \pm   4.6 $\\
\ion{C}{vi} Ly\,$\alpha$  &33.70           & n.a.          &  n.a.           &$33.70\pm0.01$  &$ 50\pm 9$     &$20   \pm3.6$&$ 73.0 \pm  56.6 $\\
\ion{Fe}{xvii}            &15.01           &$15.01\pm0.01$ & $50 \pm 9  $    &$15.03\pm0.01$  &$101\pm13$     &$11.5\pm1.7 $&$ 35.5 \pm  11.6 $\\
\ion{Fe}{xvii}            &15.26           & n.a.          & n.a.$^b   $     &$15.28\pm0.01$  &$68\pm11$      &$ 8.7\pm1.4 $&$ 26.8 \pm   9.0 $\\
\ion{Fe}{xvii}            &16.78           &$16.79\pm0.01$ & $26 \pm 7 $     &$16.78\pm0.01$  &$59\pm11$      &$6.6 \pm1.3 $&$ 20.0 \pm   7.6 $\\
\ion{Fe}{xvii}            &17.05$^d$       & n.a.          &   n.a.          &$17.05\pm0.01^c$&$53\pm14$      &$7.4 \pm1.9 $&$ 22.4 \pm   9.3 $\\
\ion{Fe}{xvii}            &17.10$^d$       & n.a.          &    n.a.         &$17.10\pm0.01^c$&$59\pm14$      &$8.2 \pm1.9 $&$ 24.9 \pm  10.0 $\\
\hline
\end{tabular}
\end{center}
$^a$ line blended (see text) \\
$^b$ region contains empty bins\\
$^c$ lines are not resolved, but two components are necessary to reproduce the line width, for the fit we fix the difference in $\lambda$
\end{table*}
Only three lines are measured in both RGS detectors. In this case we use the error-weighted mean of the RGS1 and the RGS2 flux. The photon fluxes of \ion{O}{viii} Ly\,$\alpha$ are fully consistent between the RGS1 and the RGS2 within the errors; for \ion{Fe}{xvii}~15.01\AA{} and \ion{Fe}{xvii}~16.78\AA{} the derived fluxes differ by more than the formal $1\sigma$-error due to the uncertain determination of the background level in particular in the presence of the very close \ion{Fe}{xvii}~15.26\AA{} line.

We significantly detect lines of neon, oxygen and carbon in the He-like or H-like ionisation stages and, with a significance just above $2\sigma$, nitrogen. We also find a number of \ion{Fe}{xvii} lines, which have a peak formation temperature around $8\times10^6$~K.  From the fitted line fluxes we calculate the total line intensity, using the absorption cross section from \citet{1992ApJ...400..699B}. The relative errors on the line intensity are larger than those on the line fluxes, because here the uncertainty in the fitted $N_{\mathrm{H}}$ and the distance contributes to the total error budget.

\subsection{He-like triplets}
\begin{figure}
\resizebox{\hsize}{!}{\includegraphics[angle=-90]{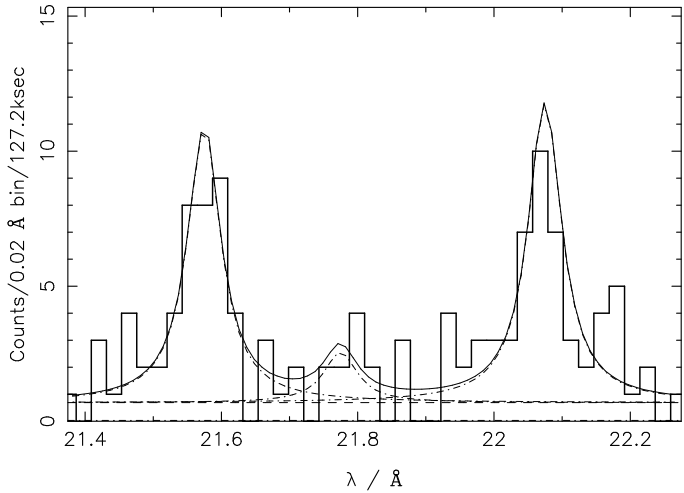}}
\caption{\ion{O}{vii} He-like triplet with our best fit to recombination, intercombination and forbidden line.}
\label{otrip}
\end{figure}

Of special interest are the line fluxes of the He-like triplets in \ion{O}{vii} and \ion{Ne}{ix}. These triplets consist of a recombination (r), an intercombination (i) and a forbidden (f) line \citep{1969MNRAS.145..241G,2001A&A...376.1113P}. So-called R- and G-ratios are defined as $R = f/i$ and $G = (f+i)/r$ respectively; for high electron densities or strong UV photon fields the $R$-ratio drops below its low-density limit, because electrons can be excited from the upper level of the forbidden to the intercombination line collisionally or radiatively. Figure~\ref{otrip} shows the \ion{O}{vii} He-like triplet in the RGS1 detector.
The $R$-ratio is $4.6$, which is above the low-density ratio of 3.4 as obtained from the CHIANTI database \citep{CHIANTII,CHIANTIVII} or 3.95 from APEC \citep{APECAPED}. 
The error on this ratio is dominated by the statistical error on the weak i line and is significantly asymmetric. We run a Monte-Carlo simulation to obtain lower limits on the $R$-ratio, which are described in detail in appendix~\ref{error}. We determine the background from the line-free regions in the range 20-25~\AA{} to 30~counts~\AA$^{-1}$. 
At a 90\% confidence level the $R$-ratio is higher than 2.6, at a 99\% confidence level the lower boundary is 1.7, it is thus fully compatible with the low-density limit, but not with a high-density case.
The $G$-ratio is $1.2_{-0.4}^{+0.6}$ (90\% confidence), where the errors are based on the same method. According to the CHIANTI database this diagnoses a plasma temperature between 0.02~keV and 0.2~keV, a range below the coolest component fitted in the global emission measure analysis, but in nature the plasma is not divided between three components of fixed temperature, instead each component has to be interpreted as a representative for a range of temperatures. \ion{O}{vii} forms only on the cool end of that distribution, above 0.20~keV \ion{O}{viii} is the dominant ionisation stage. Therefore it is not surprising to find a lower temperature from the \ion{O}{vii} $G$-ratio. A small $G$-ratio could also be due to photoexcitation \citep{2000A&AS..143..495P}.

\begin{figure}
\resizebox{\hsize}{!}{\includegraphics[angle=-90]{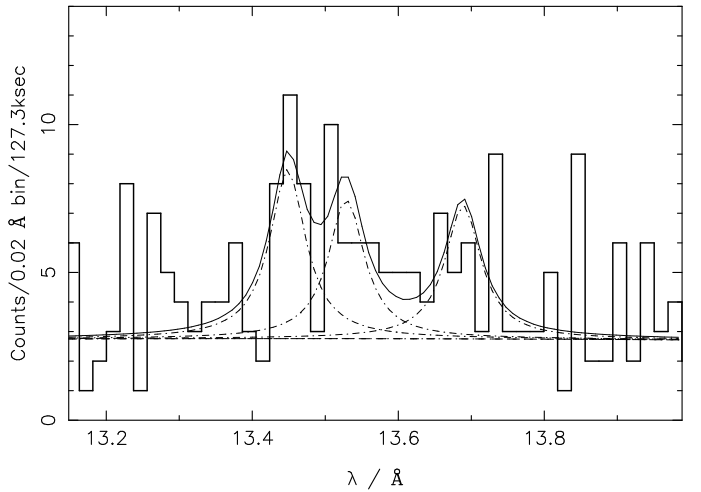}}
\caption{\ion{Ne}{ix} He-like triplet with our best fit to r, i and f line. The i lines is blended with strong iron emission.}
\label{ne9}
\end{figure}

The interpretation of the \ion{Ne}{ix} triplet is more difficult as the i line can be strongly blended by iron lines, predominantly \ion{Fe}{xix} and \ion{Fe}{xx}. The signal is not sufficient for a detailed deblending as in \citet{2008MNRAS.385.1691N}. However, there are two indications, that the counts in the i line are dominated by iron contamination: First, we fitted the He-like Ne triplet keeping the relative wavelength constant, the results of this step are given in table~\ref{lineflux} and shown in Fig.~\ref{ne9}. 
Then we kept the wavelength for the r and f line where the blending is less severe and set it free for the i line. The best fit is $\lambda=(13.53\pm0.09)$~\AA{}, just between \ion{Fe}{xix} at 13.52~\AA{} and the \ion{Ne}{ix} intercombination line at 13.55~\AA{}. Second, from a theoretical point of view, we simulated the spectra in the region of the \ion{Ne}{ix} He-like triplet with CHIANTI using the emission measures from table~\ref{3vapecem} and the abundances given in table~\ref{3vapecab}. For the spectral resolution of the RGS detectors three distinct emission peaks are expected. The middle one is always dominated by iron lines, even in the case of high densities or strong UV-fields. It is always predicted with a photon flux comparable to the r line. However, the f line is dominated by emission from neon. It is detected with the same flux as the r line in our observation (table~\ref{lineflux} and Fig.~\ref{ne9}), but should be nearly absent in the case of high densities or strong UV-fields. From these arguments we cannot obtain a numerical $f/i$ value, but qualitatively it is clear that the $R$-ratio of the \ion{Ne}{ix} triplet is high similar to the $R$-ratio of the \ion{O}{vii} triplet.

\begin{figure}
\resizebox{\hsize}{!}{\includegraphics{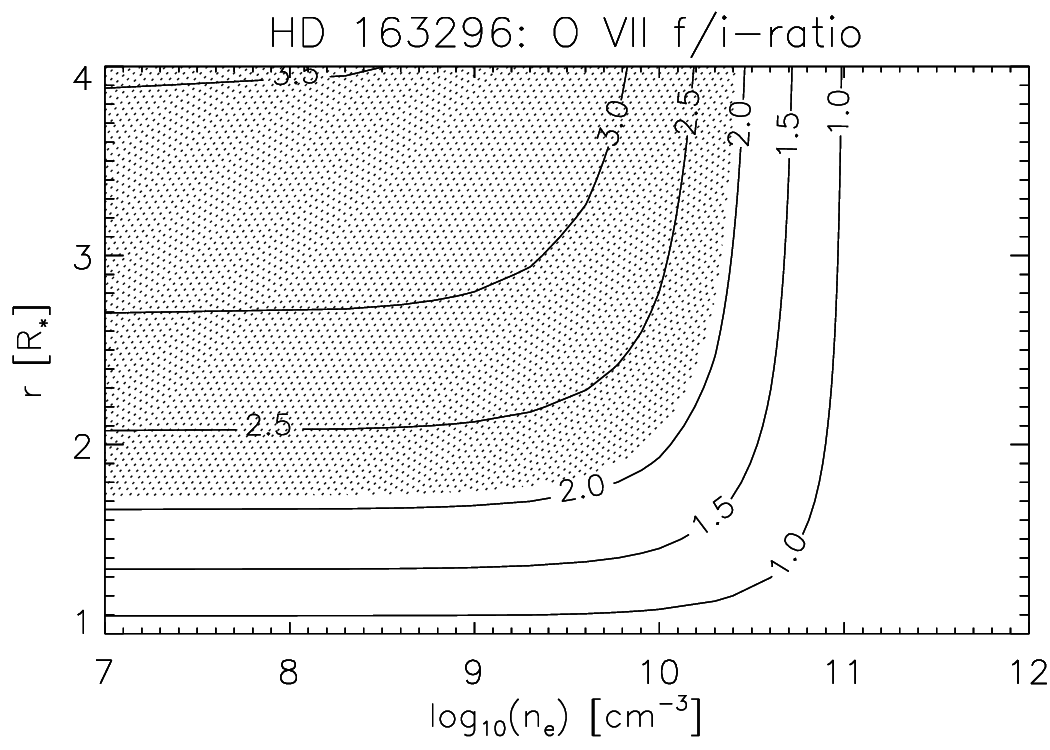}}
\caption{\ion{O}{vii} He-like triplet $f/i$ ratio depending on the density of the emitting region and the position. Shaded are those regions of the parameter space within the 90\% confidence range.}
\label{r-ratio}
\end{figure}

The theory of the He-like triplets is well developed \citep{1969MNRAS.145..241G,2001A&A...376.1113P} and the $R$ and $G$-ratios are understood. To calculate the expected $f/i$ we use the formulae given in \citet{1972ApJ...172..205B}:
\begin{equation}
R=\frac{f}{i}=\frac{R_0}{1+\Phi/\Phi_c+n_e/N_c}
\end{equation}
where $R_0$ denotes the low density limit (3.95 for the case of \ion{O}{vii}). $N_c$ and $\Phi_c$ represent the critical density and photon field respectively, where the $R$-ratio becomes density or radiation sensitive. $n_e$ is the electron density. The relevant wavelength to excite an electron from the upper level of the f to the i line is 1630~\AA{} in the \ion{O}{vii} triplet and we obtain the corresponding stellar flux from an \emph{IUE} observation \citep{2000ApJS..129..399V}. We deredden the flux with $A_V=0.25$ \citep{1998A&A...330..145V} according to the formula of \citet{1989ApJ...345..245C}. This in turn gives a radiation temperature and allows to calculate the stellar radiation field to get $\Phi/\Phi_c$ \citep[for details see][]{2002A&A...387.1032N}. We use a geometric dilution factor $W$ to express the decrease of the radiation field with increasing distance $r$ to the stellar surface $R_*$:
\begin{equation}
W=\frac{1}{2}\left(1-\sqrt{1-\left(\frac{R_*}{r}\right)^2}\right)
\end{equation}
Figure~\ref{r-ratio} shows contours of the expected $R$-ratio in the \ion{O}{vii} He-like triplet depending on the density of the emitting region and its position, expressed in stellar radii, where we define the stellar surface as $r=1$. The region compatible with the observation is shaded in grey.
Qualitatively this agrees with the requirement of a large $R$-ratio in the \ion{Ne}{ix} triplet. 
Clearly we need to look for emission scenarios, where the emission region is located above the stellar surface and the electron number density is below $10^{10}$~cm$^{-3}$.

The observation of AB~Aur is less well exposed. Additionally the i line in the \ion{O}{vii} triplet contains two missing data bins caused by damaged rows in the detector. The f line is clearly present and there can be no contamination for oxygen by the near, bright source SU~Aur because it is strongly absorbed. So we agree with \citet{ABAur} that there are indications for a large $f/i$ ratio, but we cannot give a reliable number of counts in the i line. For neon we find $f/i=2.4^{+3.4}_{-1.3}$ (90\% confidence). Due to our small extraction region the contribution of SU~Aur in the \ion{Ne}{ix} triplet should be negligible, furthermore AB~Aur seems to be iron depleted \citep{ABAur}, but any iron contamination would further enhance the $f/i$ ratio. This observation favours low-density emission regions above the stellar surface, too.

\subsection{Cool excess}
\label{coolexcess}
\begin{figure}
\resizebox{\hsize}{!}{\includegraphics{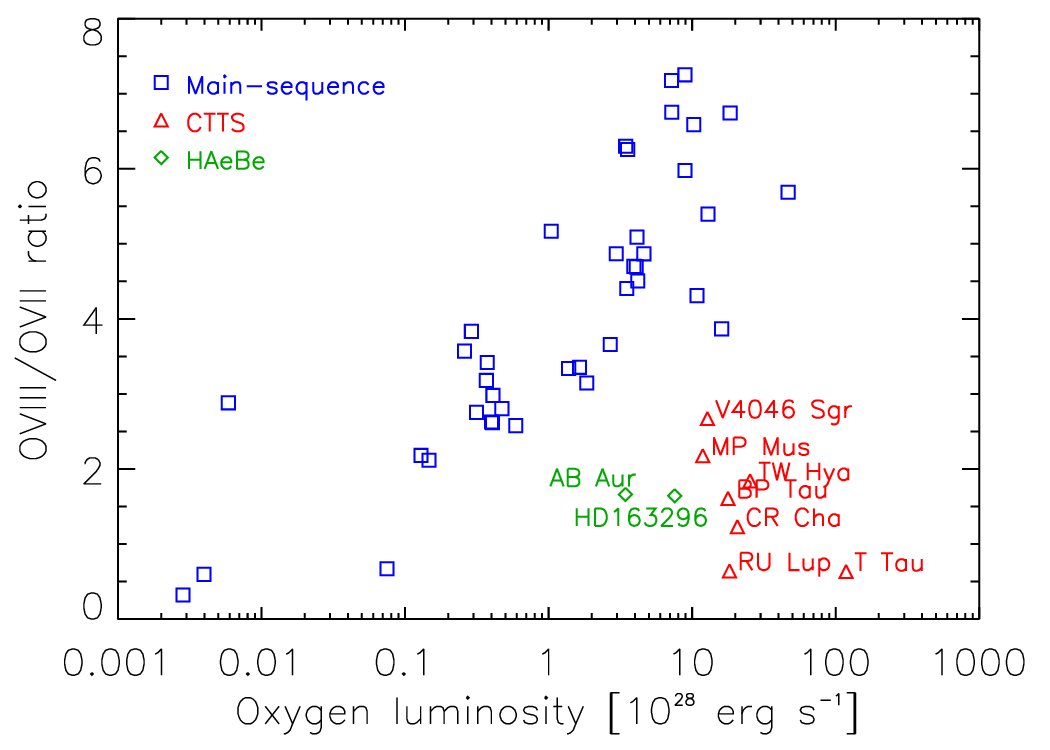}}
\caption{Ratio of \ion{O}{viii} flux to \ion{O}{vii} flux. (colour in electronic version only)}
\label{o82o7}
\end{figure}

\citet{RULup} and \citet{manuelnh} present an analysis of the \ion{O}{viii}/\ion{O}{vii} ratio as a measure of the excess of soft emission in CTTS is respect to stars on the MS. These studies show the observed excess to be confined to a narrow temperature range about the formation of the He-like \ion{O}{vii} triplet at 1-2~MK. In Fig.~\ref{o82o7} we add the two HAeBe stars with resolved \ion{O}{vii} and \ion{O}{viii} lines to their sample. The MS stars are taken from \citet{2004A&A...427..667N}. 
The sample of CTTS has larger total luminosities in oxygen than the HAeBes. This is a selection bias because the analysis requires a clear separation of the \ion{O}{viii} and \ion{O}{vii} line which only grating spectra allow. Those observations usually probe the most luminous members of each class. The \ion{O}{viii} to \ion{O}{vii} ratio for a given luminosity is smaller for all young stars than for MS stars indicating that the younger stars are cooler. Within the CTTS there is a tentative trend with the youngest stars in the Taurus molecular cloud and the Lupus-Auriga star forming region at the bottom and older objects like \object{MP Mus} and TW~Hya closer to the MS. We can plot the ratio only for two HAeBes with approximately the same age, but both are set clearly apart from the MS. As for CTTS MS stars of the same luminosity would be considerably harder. In CTTS this soft excess is presumably caused by accretion but a contribution from winds and outflows is also possible. We present our interpretation of the cool excess in HAeBes in Sect.~\ref{softcomp}.

\section{Discussion}
\label{discussion}
First, we compare our results for HD~163296 to those obtained for AB~Aur and HD~104237 (Sect.~\ref{compabaur}). We present reasons to reject the hypothesis that an as yet undetected companion is responsible for the X-ray emission in Sect.~\ref{companion}. In Sect.~\ref{accretion} we then discuss the speculation by \citet{2005ApJ...628..811S} on the basis of imaging \emph{Chandra} data, that HD~163296 may be accretion dominated like the classical T~Tauri star TW~Hya. A full review of many possible X-ray generation mechanisms including accretion, winds, and disk-related models is given in \citet{ABAur}. Because little information is available to constrain many of those models from X-ray observations we refer the reader to that publication for a larger variety of models. In Sect.~\ref{softcomp} we discuss which mechanism might be responsible for the soft X-ray emission and in Sect.~\ref{hardcomp} we present a corona from primordial fields as a good candidate for the origin of the hard component.

\subsection{Comparison to AB Aur and HD~104237}
\label{compabaur}
HD~163296 and AB~Aur share many characteristics. They are of the same age and the spectrum differs only by one or two subtypes around A0. HD~104237 is of spectral type A8. It thus seems natural to expect similar X-ray properties from these stars. And, indeed, the total luminosity of $\log L_X$ for the first two stars is 29.6~erg~s$^{-1}$, HD~104237 is brighter ($\log L_X\approx30.5$~erg~s$^{-1}$), although a contribution of about 10\% from  HD~104237-B has been removed from the grating spectrum \citep{2008ApJ...687..579T}. It is unknown how much luminosity the HAeBe primary contributes and which fraction is due to the unresolved, close CTTS companion of HD~104237. In a fit with two emission components about one third of the total emission measure is found at 0.2~keV for AB~Aur, in contrast to HD~163296 where, according to table~\ref{3vapecem}, 60\% fall in this temperature region. Due to their lower signal \citet{ABAur} do not split the hotter component as we do. Their temperature is comparable to our medium component. We find the hardest component in HD~163296 to be the one with the lowest emission measure, although it is significantly detected in our observation. If the same emission mechanism is invoked for AB~Aur and HD~163296 then a convincing explanation needs to be found why HD~163296 is softer. HD~104237 is significantly hotter, only little emission measure is found below 0.25~keV in the reconstructed emission measure distribution. Again, the hot plasma might be due to the unresolved CTTS.

As for elemental abundances we left carbon as a free parameter instead of nitrogen as \citet{ABAur} did. To compare the fitted values for O, Ne, Mg, Si and Fe we first need to correct them to the same reference values. On the one hand, in AB~Aur Ne and Si have the largest abundance and O, Mg, and Fe the smallest, in HD~163296 on the other hand C, Si, Mg and Fe, those elements which condense on grains easily, are found with enhanced abundance. A very similar pattern is found from the spectrum of HD~104237, only here S was left as a free parameter of the fit instead of C.
\begin{figure}
\resizebox{\hsize}{!}{\includegraphics{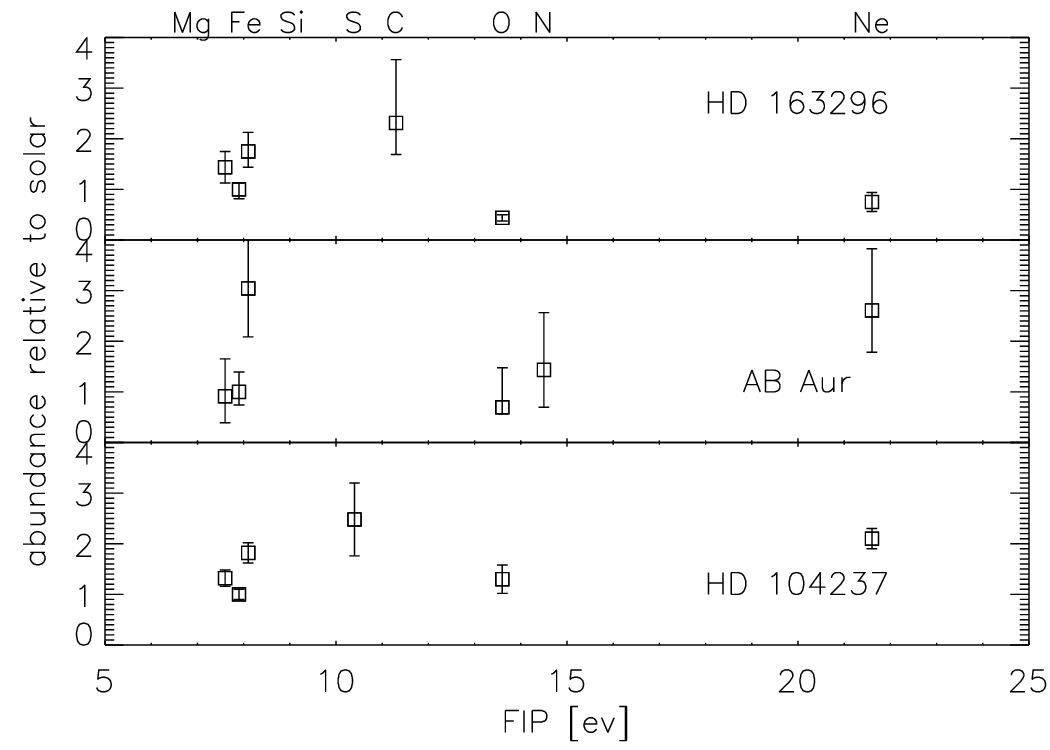}}
\caption{Abundances for HD~163296, AB~Aur \citep{ABAur} and HD~104237 \citep{2008ApJ...687..579T}. Error bars show $1\sigma$ uncertainties and all abundances are normalised to iron.}
\label{fip}
\end{figure}
We show the abundance pattern of all three stars in Fig.~\ref{fip}. The abundance distribution of AB~Aur looks erratic, whereas in HD~163296 there is trend with the FIP effect. For HD~104237 the pattern is very similar to HD~163296 except for a higher neon abundance.
However, in young objects as HD~163296 with ongoing accretion the physical mechanism might be different: The same abundance pattern can also be interpreted as an overabundance of elements released from refractory grains. The noble gas neon does not condense as easily on grains as the other elements in table~\ref{3vapecab}. 

For line emission spectra with only little continuum the absolute metallicity is very uncertain. Although the luminositys of both stars are equal, \citet{ABAur} find a larger emission measure for AB~Aur and at the same time a lower metallicity. This likely does not point to a physical difference but only shows an ambiguity in the fitting process, so we normalised all abundances in Fig.~\ref{fip} with respect to Fe. If different emission mechanisms, e.g. a corona and a jet contribute to the emission than those abundances represent an average value. Therefore we conclude than an emission mechanism for soft X-rays, which is possibly based in a region of now gaseous former refractory grains, operates in HD~163296 much stronger or in addition to the processes active in AB~Aur and possibly HD~104237. 

\subsection{A companion?}
\label{companion}
Because HAeBe stars are not expected to posses an outer convective layer, X-ray emission detected from these objects is often attributed to low-mass companions. The high optical luminosity of the main component could overwhelm the optical emission of close CTTS. In several systems high resolution \emph{Chandra} images allow to separate multiple components \citep{2006A&A...457..223S,BeateHAeBe}, still as yet undetected companions closer to the central HAeBe star might exist. In the case of HD~163296 there is no second stellar source resolved in the images and the source position matches the expected coordinates within $1\sigma$ of the \emph{Chandra} pointing accuracy \citep{2005ApJ...628..811S}. Furthermore, the same authors obtained \emph{HST}/STIS data with no sign of binarity. All this makes an origin on a late-type companion unlikely. Also, the X-ray data differs significantly from both a CTTS in the $f/i$ ratio and a weak-lined T~Tauri star (WTTS) which display activity seen in hot plasma and variable, flaring lightcurves \citep{2005ApJS..160..401P}.

\subsection{Accretion?}
\label{accretion}
The relative softness of the spectrum lead \citet{2005ApJ...628..811S} to speculate that HD~163296 may be dominated by a strong accretion component. Accretion of matter could proceed along magnetically funnelled streams from the disk to the stellar surface. This is the standard scenario for CTTS \citep{1984PASJ...36..105U,1991ApJ...370L..39K}. The accretion proceeds with free-fall velocity and a strong shock develops on the stellar surface. 
We fitted HD~163296 with an accretion shock model \citep{acc_model} and two \texttt{VAPEC} components. In short, our model takes the infall velocities and densities as input. It heats the matter up in a strong shock and follows the cooling in the post-shock accretion zone in a 1D geometry. The code explicitly takes into account non-equilibrium ionisations. The soft emission around 0.2~keV  and some of the medium emission could be produced by a shock with low densities of $4\times10^{10}$~cm$^{-3}$ or less which is falling in at around 550~km~s$^{-1}$. A filling factor $f=7\times 10^{-4}$ of the stellar surface is sufficient to produce the observed luminosity. At the given shock parameters about 20~\% of the total accretion luminosity are emitted in X-rays, the mass accretion rate in this case would be $\dot M=10^{-11} M_{\sun}$~yr$^{-1}$. Given the stellar parameters of AB~Aur a similar scenario could apply to that star, too. The accretion rate is much lower than the $\dot M=10^{-8}-10^{-7} M_{\sun}$~yr$^{-1}$ found by optical observations \citep{2006A&A...459..837G} or modelling of the spectral energy distribution (SED) \citep{2006MNRAS.365.1283D}. This resembles our results for CTTS \citep{acc_model,Bonn06} and we attribute it to non-uniform accretion spots, which only produce X-rays in the innermost hottest and densest parts of the spot. Our argument contrasts \citet{ABAur}: They take the optically determined accretion rate and the density and distribute it over the stellar surface, which results in a flux much larger than observed. Our models are more sophisticated and explicitly give the fraction of the accretion luminosity that is radiated in the soft X-ray band, but they are invalid in the case of strong absorption. The low column density found in the observation supports the validity of our model. We expect the post-shock cooling zone to penetrate the stars down to $0.97R_*$, but most of the X-ray emission is generated close to the stellar surface. Due to the low-density of the pre-shock material the optical depth is low and the X-rays escape. Observationally a large optical depth would be seen as a significantly suppressed resonance line in the He-like triplets, whereas the intercombination and the forbidden lines remain unaltered. In all stars analysed so far the strength of resonance line is compatible with the expected $G$-ratio, thus proving that the emission is optically thin.

In addition to the energetics the accretion shock model explains the small $f/i$ ratios found in CTTS with the high densities in the post-shock cooling zone \citep{acc_model}, but in HD~163296 and AB~Aur we find an $f/i$ which is fully compatible with the low-density limit. This excludes regions close to the stellar surface as emission origin because here the stellar radiation field of an A type star would radiatively shift emission from the f to the i line, if the accretion zone is not shielded from the radiation. If it is, e.g. by the outer layer of the accretion stream, then the X-ray emission should be more absorbed, but the measured absorbing column density agrees well with the stellar optical reddening (Sect.~\ref{fitem}) already without an extra absorption component. So we reject the original idea by \citet{2005ApJ...628..811S} that the soft X-ray emission in HD~163296 is powered by accretion shocks.


\subsection{The origin of the soft component}
\label{softcomp}
The soft radiation in our three component model originates in a plasma with temperatures around 0.21~keV$\approx2.4\times 10^6$~K. Plasma can be heated to this temperature if gas moves with  and passes through a strong shock (it can be slower then the 550~km~s$^{-1}$ mentioned above because we try to explain the soft component only), according to the Rankine-Hugoniot conditions, which lead to the following formula:
\begin{equation} 
\left(\frac{v_{\mathrm{shock}}}{400\textnormal{ km s}^{-1}}\right)^2 \approx 
\frac{\textnormal{k}T}{0.21\textnormal{ keV}}  ,
\label{vkt} 
\end{equation}
where we transform the kinetic energy of the pre-shock velocity $v_{\mathrm{shock}}$ in the shock rest frame into the post-shock thermal energy k$T$. Other situations than the accretion shock discussed above can give rise to such scenarios. Shocks occurring in winds or outflows easily fulfil the conditions on the density and distance to the photosphere set by the analysis of the He-like tiplets. In AB~Aur temporal variation in the X-ray luminosity was detected which is compatible with the period of 42~h, observed in \ion{Mg}{ii} and \ion{He}{i} lines which are tracers of winds and outflows \citep{ABAur}. Shocks could develop where fast and slow winds collide either through magnetic collimation or because of temporal variability of the launching velocity. 
Here we want to discuss a scenario not covered by \citet{ABAur}.

HD~163296 drives two powerful collimated jets which are visible in coronographic images obtained with \emph{HST} \citep{2000ApJ...542L.115D,2000ApJ...544..895G}. In this respect it is similar to the lower mass CTTS DG~Tau, where also X-ray emission has been detected from the jets \citep{2005ApJ...626L..53G,2008A&A...478..797G}. The high extinction of the central source DG~Tau itself allows to detect a significant offset between the soft and the hard X-ray emission \citep{Schneider}, because it absorbs all contributions from the star in the low energy band. This offset points to an origin of the soft emission in the jet forming region, possibly as a collimation shock or in an emerging knot where faster ejecta catch up with previous older outflows \citep{dgtau}. We searched for an offset in the \emph{Chandra} data for HD~163296, but could not find any deviation from the expected PSF, so the emission region must be within 100~AU of the star. Line ratios of \ion{Fe}{ii} indicate electron densities between $10^4$ and $10^5$~cm$^{-3}$ in the knots of the jet \citep{2006ApJ...650..985W}. Presumably the density increases towards the star, so we assume a pre-shock particle number density of $n_0=10^6$~cm$^{-3}$ in the shock region. With this assumption we can apply the approximation to the post-shock cooling length $d_{\mathrm{cool}}$ of \citet{2002ApJ...576L.149R}:
\begin{equation}
d_{\mathrm{cool}} \approx 0.8 \mathrm{ AU} 
    \left(\frac{10^6\mathrm{ cm}^{-3}}{n_0}\right) 
    \left(\frac{v_{\mathrm{shock}}}{400\textnormal{ km s}^{-1}}\right)^{4.5}
\label{raga} 
\end{equation}
and find that $d_{\mathrm{cool}}$ is only 0.8~AU. 
Dividing the volume emission measure (VEM) by the density we obtain the volume and - using the equation above - we can transform this to a radius $R$ of the shock front (in AU) \citep[for details see][]{dgtau}
\begin{equation}
R \approx 0.5 
        \left(\frac{10^6 \mathrm{cm}^{-3}}{n}\right)^{0.5}
        \left(\frac{VEM}{10^{52} \textnormal{cm}^{-3}}\right)^{0.5} 
        \left(\frac{0.21 \mathrm{keV}}{\mathrm{k}T}\right)^{1.125},
\label{eqnr}
\end{equation}
where we assume a cylindrical geometry with the shock at the cylinder base. The mass flux $\dot M_{\rm shock}$ through the shock is independent of the density, because smaller densities lead according to Eqn.~\ref{eqnr} to larger shock areas:
\begin{equation}
\dot M_{\rm shock} \approx 6\cdot 10^{-11}\frac{M_{\sun}}{\rm yr} 
              \left(\frac{VEM}{10^{52}\textnormal{ cm}^{-3}}\right)
              \left(\frac{0.21\textnormal{ keV}}{\textnormal{k}T}\right)^{1.75} 
\label{mdot1}
\end{equation}
For HD~163296 a mass outflow of the order $\dot M_{\rm shock} \approx 10^{-10} M_{\sun}$~yr$^{-1}$ is sufficient to power the soft X-ray component, this is two orders of magnitude smaller than the mass loss rates determined from UV and optical data \citep{2005A&A...429..247D,2006ApJ...650..985W}. Only a small component of the total jet needs to be fast enough to be shocked to X-ray emitting temperatures.

This scenario seems feasible in the light of the optical observations. The velocity of the jet determined from a combination of radial velocity and proper motion is $v_{\rm jet}=360\pm82$~km~s$^{-1}$ \citep{2006ApJ...650..985W} where the radial velocity is taken as the average of a broader distribution at the position of the innermost knot. The offset between the central source and the knot A is 9\arcsec{}, at the distance of HD~163296 this corresponds to 1100~AU. Certainly the jet is launched faster and decelerates as it goes through the shock and later moves through the surrounding matter. In the shock rest frame the velocity $v_1$ drops to a quarter of the pre-shock velocity $v_0$, where the matter passes through the shock front: $4\cdot v_1=v_0$. To heat the plasma to the required temperatures we need $\Delta v=v_0-v_1=400$~km~s$^{-1}$. This is all fullfilled if the shock front travels outward along the jet at $v_{\rm front}=230$~km~s$^{-1}$ in the jet rest frame. The jet emerges at $v_{\rm launch}=v_{\rm front}+v_0=750$~km~s$^{-1}$ and the knot moves at $v_{\rm jet}=v_{\rm front}+v_1=360$~km~s$^{-1}$ with respect to the star.

Typical jets of CTTS reach up to 400~km~s$^{-1}$ \citep{1998AJ....115.1554E}, comparable to the case of HD~163296. \citet{1995RMxAC...1....1R} report velocities of up to 1000~km~s$^{-1}$ out in the jet, so an initial velocity of 750~km~s$^{-1}$ is not unreasonable. This X-ray generating mechanism may operate in addition to any of those already suggested for AB~Aur or HD~104237. Possibly HD~163296 is softer because in AB~Aur has no jet which could emit soft X-rays.

Several models for the acceleration and collimation of outflows have been suggested even for the class of CTTS where the sample size is much larger than for HAeBes, still the exact collimation mechanism is uncertain. Yet, all models rely on a magnetic field.
So the existence of two jets already makes the presence of magnetic fields in the environment of HD~163296 very likely. Spectropolarimetric observations by \citet{2007A&A...463.1039H} support this idea with signatures of magnetic fields in \ion{Ca}{ii} H and K lines likely of circumstellar origin. Unfortunately the signal does not allow a qualitative measurement of the field strength which is needed to estimate the radius of the magnetosphere for alternative models of magnetically confined winds. Also \citet{2007MNRAS.376.1145W} have questioned the method used in theses studies.

\subsection{The origin of the hard component}
\label{hardcomp}
To heat the plasma emitting the hot component shocks with velocities above 1300~km~s$^{-1}$ are needed but the optical observations show no lines which are shifted by this amount. Other heating mechanism need to be responsible and a promising candidate are reconnection events in magnetic fields in analogy to the solar and other late-type coronae. No  polarisation is found in photospheric lines \citep{2006A&A...446.1089H,2007A&A...463.1039H} but this may be due to strong small-scale fields with only little global dipole contribution. Young A stars may either still posses primordial magnetic fields or generate there own. In this evolutionary stage the simulations of \citet{2000A&A...358..593S} predict HAeBes to have a thin outer convective layer where a dynamo could operate or the shear in rapidly rotating stars could cause magnetic fields \citep{1995MNRAS.272..528T}. Given the detection of polarisation in circumstellar matter and the collimation of the outflow we expect that magnetic fields exist on HD~163296 itself.

No flaring activity has been unambiguously observed on a HAeBe star and especially AB~Aur, HD~104237 and HD~163296 show only little variability in their hard components during the observation, but \citet{2004A&A...413..669G} can at least give good indications that the flare they observed occurs on a HAeBe star and not on the known, but unresolved CTTS companion. 
We postulate that the medium and hard component of the X-ray emission originate in a stellar corona. At high temperatures the contribution to the \ion{O}{vii} and the \ion{Ne}{ix} He-like triplet is small so the observed $f/i$ ratio cannot trace the UV field in the emission region of the hard photons.

\section{Conclusion}
\label{conclusion}
We present the observation of HD~163296 with \emph{XMM-Newton}. This is after AB~Aur the second grating spectrum of a HAeBe, without -as far as we know- an unresolved companion. The signal-to-noise ratio (SNR) is significantly better than in the first observation. Congruently both stars show soft spectra with little temporal variability. In all ana\-lysed He-like triplets the line ratios point to an emission origin in regions of low densities and weak UV photon field, that is at a few stellar radii. On HD~163296 we find a FIP abundance pattern typical for inactive stars. 

Both the spectral shape and data from other instruments make an origin of the X-ray in an unknown companion to HS~163296 very unlikely. Therefore we believe to present genuine X-ray data for HAeBe which can be representative for the whole class of this objects for all we know today.

The optically observed mass flux rates are much larger than required by the energetics of the X-ray radiation in the case of accretion shocks or emission from outflows, but the large $f/i$-ratios rule out an accretion shock scenario on the stellar surface. We favour a model where the main contribution to the soft X-ray emission comes from shock heated plasma in the jet or wind.  Possible geometries include shocks in unsteady winds or collimation shocks on the base of the jet. 
The collimation of the jets is likely related to a magnetic field, whose origin is unclear. This field can be a frozen-in primordial magnetic field or the star generates one in a thin convective layer. We find no other convincing explanation for the hard emission than heating by energetic events from the magnetic field. Thus we predict that with longer observation times we should observe flare-like behaviour on HAeBes. 

\begin{acknowledgements}
H.M.G. acknowledges support from DLR under 50OR0105.
\end{acknowledgements}

\begin{appendix}
\section{Error estimation for line ratios}
\label{error}
For experiments with low count numbers the observed counts can be taken to be Poisson-distributed. For longer exposure times or stronger sources the Poisson distribution can be reasonably approximated by a Gaussian distribution. In this case the error on a line ratio can be calculated according to the conventional error propagation formulas. For weak sources, however, the Poisson distribution leads to significantly asymmetric errors, which are dominated by the statistical uncertainty in the weakest line.

We set up a Monte-Carlo simulation to determine confidence intervals. The instrumental and source background is combined in a total background $b$, which we assume to be uniform over a range of a few \AA. It is measured from a larger line-free region on both sides of the triplet of interest. Thus the mean background level is well known. In the case of the \ion{O}{vii} triplet in HD~163296 we determine the background from the line-free regions in the range 20-25~\AA{} to $b=30$~counts~\AA$^{-1}$.

For a fit with $i$ lines, the spectrum is divided into $i$ bins centred on the lines, each bin is $\Delta\lambda_i$ wide, thus the expected background per bin is 
\begin{equation}
B_i=b \cdot \Delta\lambda_i
\end{equation}

\begin{figure}
\resizebox{\hsize}{!}{\includegraphics{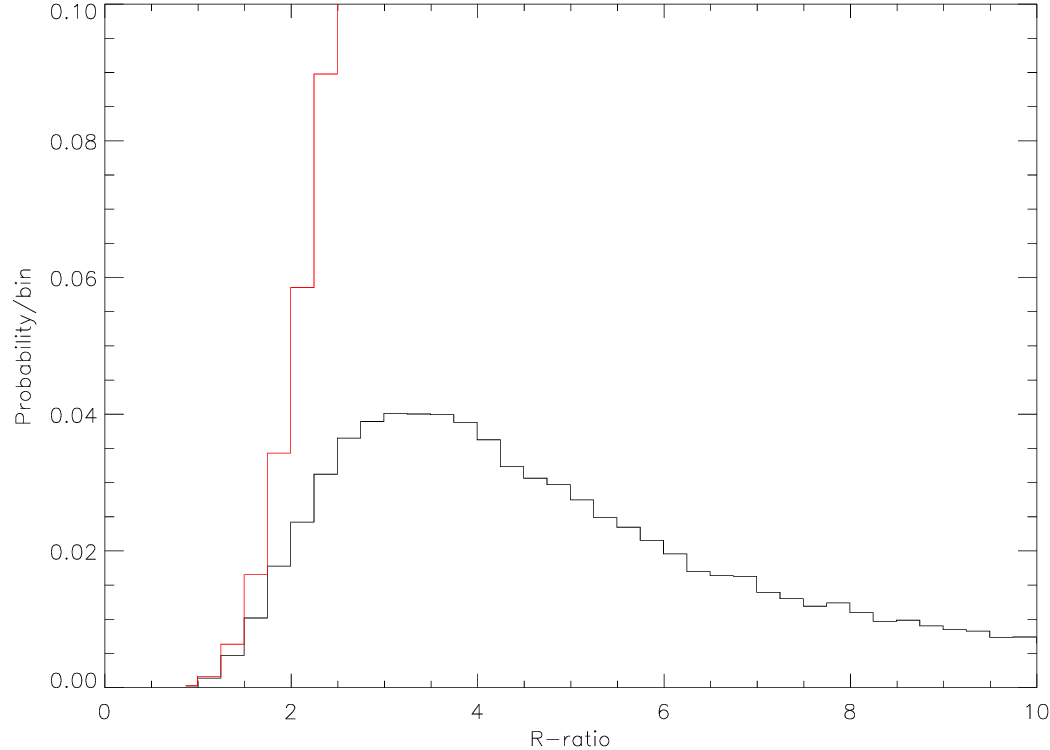}}
\caption{Statistical distribution of the $R$-ratio based on $10^5$ realisations of a Poisson statistic for the forbidden and intercombination line. The red/grey line shows the cumulative distribution function.}
\label{mcsim}
\end{figure}

The line profile is given by the instrumental profile, in the case of \emph{XMM-Newton} the data is best fit by a Lorentz-profile. For each bin $i$ we compute the fraction $a_{ij}$ of the total line spread function of line $j$ that is expected in bin $i$. For known line fluxes $f_j$ then the observed number of counts in each bin can be calculated. We write this in form of a matrix equation
\begin{equation}\label{eqn1}
\vec{L} = \tens{A}\cdot\vec{F}+\vec{B} ,
\end{equation}
where $\tens{A}$ is constructed from the $a_{ij}$, $\vec{B}$ from $B_i$ and $F$ from $f_j$. $\vec{L}$ represents the observed counts per spectral region. Since the sum of Poisson distributions is again Poisson-distributed, we can take the number of counts in each bin as Poisson-distributed, even it is calculated as the sum over several detector bins. Thus the values of $\vec{L}$ are Poisson-distributed and we simulated 100\,000 realisations of a Monte-Carlo calculation for each bin, taking the observed count number as expectation value. Inverting Eqn.~\ref{eqn1} gives
\begin{equation}
\vec{F}=\tens{A}^{-1}\cdot\left(\vec{L}-\vec{B}\right).
\end{equation}
From the simulations we use this to obtain the values of the individual line fluxes $f$. Negative numbers occur, if the simulated value of the Poisson distribution happens to be below the fixed background value. We reset these numbers to 0; in the analyses of real data they represent situations, where the line seems to be absent, thus producing extreme values for the line ratio. The simulated count ratios are scaled by the effective area and the flux ratios are calculated. The simulated distribution of the $f/i$-ratio is shown in Fig.~\ref{mcsim}. From the cumulative distribution (red/grey line) the confidence limits can be read off.

\end{appendix}

\bibliographystyle{aa} 
\bibliography{../articles}
\end{document}